\documentclass{bmcart}

\usepackage[utf8]{inputenc} 
\usepackage{graphicx} 
\usepackage{caption}
\usepackage{hyperref}
\usepackage{subcaption}
\usepackage{tablefootnote}
\usepackage{longtable}
\usepackage{multirow}
\usepackage{array}
\usepackage{float}
\usepackage[title]{appendix}

\newcolumntype{L}[1]{>{\raggedright\let\newline\\\arraybackslash\hspace{0pt}}m{#1}}
\newcolumntype{C}[1]{>{\centering\let\newline\\\arraybackslash\hspace{0pt}}m{#1}}
\newcolumntype{R}[1]{>{\raggedleft\let\newline\\\arraybackslash\hspace{0pt}}m{#1}}

\makeatletter
\let\c@author\relax
\makeatother
\usepackage[natbib=false,sorting=nyt,style=numeric, maxbibnames=99]{biblatex}
\bibliography{bmc_article}   
\usepackage{biblatex2bibitem}

\startlocaldefs
\endlocaldefs

\begin{document}
\textcolor{red}{Accepted to appear in the International Journal of Educational Technology in Higher Education}
\begin{frontmatter}

\begin{fmbox}
\dochead{Research}


\title{Impact of combining human and analytics feedback on students' engagement with, and performance in, reflective writing tasks}


\author[
   addressref={aff1},                   
   corref={aff1},                       
   email={wannapon.suraworachet.20@ucl.ac.uk}   
]{\inits{WS}\fnm{Wannapon} \snm{Suraworachet}}
\author[
   addressref={aff1},
   email={qtnvqz3@ucl.ac.uk}
]{\inits{QZ}\fnm{Qi} \snm{Zhou}}

\author[
   addressref={aff1},
   email={m.cukurova@ucl.ac.uk}
]{\inits{MC}\fnm{Mutlu} \snm{Cukurova}}

\address[id=aff1]{
  \orgname{UCL Knowledge Lab, IOE, UCL's Faculty of Education and Society, University College London}, 
  \street{23-29 Emerald St},                     %
  \city{London},                              
  \cny{UK}                                    
}





\end{fmbox}


\begin{abstractbox}

\begin{abstract} 
Reflective writing is part of many higher education courses across the globe. It is often considered a challenging task for students as it requires self-regulated learning skills to appropriately plan, timely engage and deeply reflect on learning experiences. Despite an advance in writing analytics and the pervasiveness of human feedback aimed to support student reflections, little is known about how to integrate feedback from humans and analytics to improve students' learning engagement and performance in reflective writing tasks. This study proposes a personalised behavioural feedback intervention based on students' writing engagement analytics utilising time-series analysis of digital traces from a ubiquitous online word processing platform. In a semester-long experimental study involving 81 postgraduate students, its impact on learning engagement and performance was studied. The results showed that the intervention cohort engaged statistically significantly more in their reflective writing task after receiving the combined feedback compared to the control cohort which only received human feedback on their reflective writing content. Further analyses revealed that the intervention cohort reflected more regularly at the weekly level, the regularity of weekly reflection led to better performance grades, and the impact on students with low self-regulated learning skills was higher. This study emphasizes the powerful benefits of implementing combined feedback approaches in which the strengths of analytics and human feedback are synthesized to improve student engagement and performance. Further research should explore the long-term sustainability of the observed effects and their validity in other contexts.
\end{abstract}




\end{abstractbox}
%

\end{frontmatter}



\section*{Introduction}

Reflective writing refers to students' written journal of learning experiences over time to develop self-awareness of their own learning \cite{thorpe_reflective_2004}. It is often utilised in higher education to stimulate their thoughtful reflection on the incidents and hence promote transformative learning and practice \cite{ryan_pedagogical_2013}. Several studies showed that incorporating reflective writing could increase students' content comprehension \cite{strong_making_2001}, 
academic performance \cite{connor-greene_making_2000} and life-long learning skills \cite{boutet_evaluating_2017, thorpe_reflective_2004}. However, reflective writing is a complicated and demanding task that requires learners to be able to regulate their own learning \cite{zimmerman_becoming_1997}. Self-regulation skills allow a person to strategically regulate their behaviours and environment towards their goals \cite{zimmerman_social_1989}. To be more specific, reflective writers are required to set goals on the coverage contents and deploy multiple cognitive processes including planning, writing and revising to complete a writing task \cite{graham_role_1994}. 

There is strong evidence that pedagogically incorporating reflective writing tasks without appropriate support is unlikely to lead to effective learning due to students' challenges in engaging with independent learning \cite{cukurova_students_2018} and with critical reflective skills that can transfer into other contexts \cite{mcintosh_action_2010}. Therefore, feedback is a necessary condition to help learners engage in their writing tasks and also develop key skills associated with reflective writing practice \cite{sadler_beyond_2010}. Written feedback provided by teachers has long been recognised as an effective method to improve students’ performance \cite{page_teacher_1958}, especially in written assignments \cite{stewart_teacher_1976}. Yet, research on reflective writing has so far mainly focused on teachers' written feedback that emphasises cognitive development and content acquisition \cite{thorpe_reflective_2004, aronson_comparison_2012}, frequently overlooking other factors that also contribute to an improvement in learning, such as students' emotions, motivations and behaviours. The quality and meaningful feedback elements suggested for reflective writing tasks at the motivational and behavioural levels \cite{aronson_comparison_2012,dekker_which_2013} as well as reactions to the tone of feedback \cite{dekker_which_2013,rozental_medical_2021} are understudied.

In recent years, learning analytics provide opportunities to support students with meaningful feedback on motivational and behavioural aspects of their learning using data from the digital traces of student activities. For example, dashboards containing information about behavioural learning engagement with the online learning management platforms (e.g. resource uses, time spent and performance level) \cite{bodily_trends_2017} or those that aim to monitor and support students' motivational goals \cite{jivet2021quantum} have been designed and used to provide support for students’ learning. Even though many studies attempted to provide feedback for the reflective writing tasks, most focused on analysing the writing contents i.e. semantic complexity to provide feedback on the content of students' reflective writings \cite{shibani_constructing_2020, shibani_contextualizable_2019}. However, trace data available in digital writing platforms are rarely combined with human feedback in interventions. It is important to study the combined feedback approaches' impact on students' learning and engagement since analytics feedback and human feedback tend to have different strengths and weaknesses \cite{luckin2018machine, cukurova2019learning}. Moreover, the real-world implementations of the interventions with the analytics feedback on students' reflective behaviours and more specifically academic performance are limited. We argue that behavioural engagement feedback generated with learning analytics when combined with traditional content feedback from human educators, can lead to better engagement with, and performance in, students' reflective writing. This paper presents the results of a semester-long intervention study that investigated the combined effects of personalised behavioural analytics feedback, based on a time series analysis of digital traces from a ubiquitous online word processing platform, and human educators' feedback on reflective writing.

\section*{Literature review: Reflective Writing Support with Learning Analytics}
\medskip
Multiple research studies aim to unravel writing processes and the quality of reflective writing from different perspectives to support reflection. On the one hand, several studies in Writing Analytics utilise methodological advances in natural language processing (NLP) to analyse the contents of writing from a final learning artefact and explore predictive features for identifying writing performance, hence helping to automate reflective writing scoring \cite{shum_critical_2016}. For example, it is found that linguistic features extracted from multi-source argumentative writing essays could determine individual differences in vocabulary score and hence help develop personalised learning systems for writing \cite{oncel_automatic_2021}. Similarly, linguistic features, such as word length, sentence length, and sentence structure, could also provide valuable evidence for essay scoring in a 30-min writing task \cite{bridgeman_design_2017}. However, the model's value to predict students' performance in real-world writing tasks was very limited. This is in line with the study of Kovanovic et. al. \cite{kovanovic_understand_2018} which explored linguistic features in relation to model reflective elements namely observation, motive, feedback and goal from reflective writing documents. The authors highlighted the major drawback of the approach to be limited to the contexts studied. Thus, whilst shown to be useful for particular contexts, the reliability, validity, and cross-context generalisability of such models of writing analytics are often critiqued \cite{kovanovic_understand_2018,neto_automatic_2021,crossley_automated_2019} and still considered inappropriate for real-world implementations. 

Apart from investigating writing content for scoring purposes, some studies tried to gain insight into the process of writing aiming to support effective writing behaviours in general. In this stream of work, researchers looked into writing artefacts to potentially detect different cognitive operations involved in the process. For instance, Winograd and colleagues \cite{winograd_detecting_2021} presented an NLP approach to identify the depth of scientific reasoning in students' written work. Other studies extracted digital traces of digital learning platforms to generate a visualisation of writing processes for improving awareness and hence increasing learning performance. To illustrate, Shibani \cite{shibani_constructing_2020} presented a technique for visualising the revision behaviours by generating Automated Revision Graphs (ARG) which could provide information about the writing process and student interactions with feedback. Another work by Turkay, Seaton and Ang \cite{turkay_itero_2018} developed a writing analytics tool called \textit{Itero}, aiming at visualising temporal writing processes gathered from revision logs to promote students' self-efficacy. However, these studies only focused on the revision behaviours of students, while other writing behaviours were not considered. In addition, analytics on the writing quality is often considered limited and lacking the expected level of semantic complexity to support student reflections as a standalone solution. For instance, Gibson and colleagues \cite{gibson_reflective_2017} who provided a conceptual framework for reflective writing with an automated approach to model writing and provide feedback to students, showed the results that most participants reported helpfulness and expressed their willingness to use the tool in the future. However, as the authors stressed, the generalisation of the system was a major limitation. Since the writing analytics system was developed based on the content of the writing, the transferability of the system to other contexts and different writing contents was problematic. The issue was also highlighted in the recent work of Liu, Kitto, and Shum \cite{liu_combining_2021} as the shortcomings of the NLP-based Capability for Written Reflection (CWRef) model to capture context-dependent reflective elements and adapt for variation in specific learning design and assessment were discussed at length. Moreover, this study only proposed a model, yet hasn't incorporated it into any real-world interventions for evaluation. 

Moreover, there is a lack of literature focusing on the impact of long-term analytics interventions and an abundance of short-period and one-off intervention studies. For example, Cotos and colleagues \cite{cotos_understanding_2020} evaluated the impact of the Research Writing Tutor (RWT), a web-based tool for academic writing which can provide different levels of automatic feedback. By analysing students' revision logs, captured screens, and stimulated recalls, the authors argued that RWT can promote students' close and deliberate examination of their produced text through different types and forms of feedback. However, RWT was only used in one class period. As the authors stated, this limited time was likely to affect the evaluation of the tool. Longitudinal data from a longer period would provide more information about how RWT might improve students' writing performance. In another study by Shibani and colleagues \cite{shibani_contextualizable_2019}, they developed an automatic writing analytics tool, called AcaWriter. The tool analyses the rhetorical moves by using NLP and provides formative feedback. It was implemented in two different contexts and has been shown to have the potential of providing meaningful contextualised support for writing. However, given the fact that both of these two implementations focused on one-off tasks, whether students can migrate the skills to future writing to reap long-term benefits remains an open question.  

In addition, even if the required technical issues are resolved with regards to generating analytics of writing content to support reflective writing, considering content feedback alone might not lead to the expected learning outcomes. For example in Wingate's \cite{wingate_impact_2010} intervention study on content feedback presented, while some students improved their writing quality over time based on the suggested feedback, many other students did not. Those students pointed out that the content feedback alone has lessened their motivation and self-efficacy which resulted in their disengagement behaviours with the feedback. A similar phenomenon was observed in Mitchell, McMillan and Rabbani's study \cite{mitchell_exploration_2019} that students with low self-efficacy and high anxiety levels, experienced low capability as writers from the content feedback alone. Hence, apart from feedback on writing content, feedback on other factors such as students' behavioural engagement might be a prerequisite for effective learning outcomes. Behavioural engagement feedback appraises and supports students' commitment to their efforts toward their own learning and has been shown to support learning outcomes \cite{vytasek_analytics_2020}. For instance, previous research has shown that highly self-regulated learners develop systematic engagement patterns in reflective writing tasks which correlate with higher reflective writing performance \cite{suraworachet_examining_2021}. Similarly, several other studies reported positive effects of engagement feedback in other contexts. For example, Plak et. al \cite{plak_raising_2022} deployed behavioural feedback in a form of email nudges which could promote higher engagement in online practice exams. Iraj and colleagues \cite{iraj_understanding_2020} reported a study on the feedback concerning students' online participation, directing them towards quizzes and reviewing processes and found that there was a link between timely engagement with the feedback and success in learning. In addition, Nelson, et al. \cite{nelson_good_2012} claimed year-long persistent engagement of at-risk students, after implementing the continuing behavioural engagement feedback. These studies indicate that the analytics on writing content and the semantics of reflections might currently be considered limited in real-world implementations. However, the analytics on the behavioural aspects of students' writing might still bring significant value to their reflective writing performance. Based on this premise, our hypothesis is that behavioural engagement feedback on students' reflective writing, not as a replacement for writing content feedback but as a supplement to  it \cite{cukurova2019artificial}, can create opportunities to increase students' overall engagement with their reflective writing tasks and can increase their performance. Currently, there is a lack of studies investigating such relationships between different types of reflective writing support with analytics and students' performance in long-term interventions. In this paper, we aim to fill in this gap with a semester-long real-world study investigating the impact of an intervention that combines human educator feedback on content with students' writing engagement feedback with analytics on their reflective writing performance.  

\section*{Research questions and hypotheses}

Behavioural feedback on students' reflective writing engagement was generated from log data of their actions in a ubiquitous online word processing platform (Google Docs). Since log data can be obtained from a pervasively accessible digital writing platform and requires no context-specific sense-making process like NLP-based content analytics, the behavioural analytics generated has the potential to be generalised into other contexts. With regards to the design of the formative feedback, we adopted Hattie and Timperley's \cite{hattie_power_2007} feedback model consisting of three components for effective feedback: 1) learning goals (Where am I going?), 2) learning progress (How am I going?) and 3) activities leading to better progress (Where to next?), into our behavioural engagement feedback. We studied students’ writing engagement behaviours from two authentic higher education cohorts comparing the intervention cohort that received the additional behavioural engagement feedback and the control cohort that received no behavioural engagement feedback. Students' writing behaviours were modelled to identify differences between the high-performance and highly self-regulated students, and their peers on the other ends of these spectra. More specifically, we investigated three research questions (RQ). 

\begin{itemize}
  \item RQ1: How does the intervention based on feedback about students' reflective writing behaviours affect their engagement with the reflective writing task?
  \item RQ2: How does the feedback intervention affect the writing task engagement of students with different self-regulated learning (SRL) competence?
  \item RQ3: What are the relationships between students' writing engagement behaviours, and their final grades?
\end{itemize}

Driven by the literature reviewed above, for the first research question (RQ1), we hypothesised that behavioural engagement feedback could help promote persistent or higher writing engagement in the intervention group than in the control group. In other words, there would be significant differences in terms of quantity, weekly engagement patterns, or both, between the control and the intervention groups after the feedback intervention due to engagement encouragement from the feedback \cite{vytasek_analytics_2020}.

In relation to the second research question (RQ2), it was hypothesised that students in the intervention cohort, regardless of their SRL levels, would show persistent writing engagement after the intervention period. However, the low SRL score from the intervention cohort may particularly benefit more from the behavioural feedback by continuing to engage with the writing task after receiving the feedback compared to their engagement of the period before. Based on previous research that suggested that content-only feedback might cause high anxiety and low self-efficacy, especially on students with low SRL \cite{mitchell_exploration_2019}, it was hypothesised that the behavioural engagement feedback may help them realise the necessity to persistently work on the task; hence support engagement patterns with the reflective writing task. 

Finally, it was hypothesised that there will be a significant correlation between the engagement analytics used and students' academic performance measured by their reflective scores (RQ3). More specifically, the higher the engagement of students with the reflection task, the higher their performance was expected to be.

\section*{Methodology}

\subsection*{Educational Context}

The study was conducted within a postgraduate course for two consecutive years. Ethical approval was received through the institutional processes. All students were informed about the study with a clear information sheet and provided their written consent at the beginning of the module. In total, 81 students consented to participate: 40 in the control group (the former year) and 41 in the intervention group (the latter year). According to the demographic data, the two groups show no difference in their age range, \emph{x\textsuperscript{2}} (3, N = 81) = 4.450, \emph{p} = .217, gender, \emph{x\textsuperscript{2}} (2, N = 81) = 1.109, \emph{p} = .574, mode of study (full-time vs part-time), \emph{x\textsuperscript{2}} (1, N = 81) = .617, \emph{p} = .432, background of study, \emph{x\textsuperscript{2}} (2, N = 81) = .245, \emph{p} = .885, and working experiences, \emph{x\textsuperscript{2}} (2, N = 81) = 4.746, \emph{p} = .093. They were assigned into small groups of 4 or 5 students with mixed-gender and interdisciplinary backgrounds. Over a 10-week course, students were introduced to the topics of the design and use of educational technology and were asked to work collaboratively to propose a technical solution for an educational challenge they have chosen. For both years, the lectures were on Tuesdays. Each week, before the lectures, students were expected to 1) complete their weekly readings, 2) study pre-recorded videos, 3) participate in a cohort-level debate, and after the lectures, 4) write a weekly individual reflection on their learning experiences via a single Google Docs, before the next week starts.

This study focused on the individual reflective writing task. There were nine weeks in total for this task since the writing task was optional for the first week which was an introductory week for students to get used to the format and tools used in the module. Individual reflections were taken into account as 40\% of the student's final grade of the course. For both years, content-based formative feedback was manually provided by the tutors as in-text comments at mid-term (week 6), and summative feedback of the final grade was provided four weeks after submissions. 

\subsection*{The intervention design}

In both cohorts, the content, delivery and reflection feedback provided by the tutors were the same. However, in the intervention group, students were sent additional personalised writing engagement feedback via email apart from their content-based formative feedback on week 6 to better support them with their engagement in the writing task. As discussed in Hattie and Timperley's \cite{hattie_power_2007} effective feedback model, the personalised behavioural engagement feedback in this study consists of 1) a recap of findings from the prior study on how high SRL students behaved and how this affected their performance (learning goals), 2) a description of an individual student's writing engagement extracted from the digital writing platform (learning progress) and 3) suggestions on how to improve their own engagement (prospective activities). To be more specific, the email feedback attached a graph showing an engagement comparison across the first five weeks between individual students, the control cohort, and the intervention cohort. The feedback first introduced the engagement comparison between the control and the intervention cohorts. Then, based on the individual student's pattern of writing behaviours, the feedback suggested students to reflect regularly every week and keep up with their reflective writing tasks. Meanwhile, it also stressed that a higher number of edited contents did not necessarily lead to better learning outcomes, but a systematic pattern of reflective writing did. \nameref{appendix:email} shows the template of this email feedback.

\subsection*{Data Collection Tools}

Log data from Google Docs\footnote{\url{http://docs.google.com}} were collected and exported through a modified Google Chrome plug-in called Draftback\footnote{ \url{https://chrome.google.com/webstore/detail/draftback/nnajoiemfpldioamchanognpjmocgkbg}}. Google stores log data as revisions and the revision number represents a unique chronological auto-incremental number of the edited document. Draftback retrieves data from Google API to generate statistical summaries and visualisations of these log data to represent students' writing activities. Data prepared by Draftback is composed of the (1) \emph{type of activity} which can either be insertion or deletion, (2) \emph{starting index} within the document where a particular edited activity happens, (3) \emph{ending index} within the document where a particular edited activity ends, (4) \emph{string} or the contents have been inserted, (5) \emph{revision number} as generated from Google Docs, (6) \emph{user ID} and (7) \emph{timestamp}. We modified Draftback to be able to export log data in .csv format for further investigation of the logged data with learning analytics.

\subsection*{SRL instruments and clustering}

In order to investigate participants' SRL competence, students from both years were asked to fill in the same standardised self-report questionnaire at the beginning of the module. The questionnaire was adapted from a meta-analysis of SRL \cite{sitzmann_meta-analysis_2011} consisting of four main dimensions, namely goal-setting (GS, 4 items), effort (E, 2 items), self-efficacy (SE, 9 items), and persistence (P, 10 items), that had the strongest effects on students' academic performance. The adapted version of the questionnaire for the studied contexts can be found in \nameref{appendix:questionnaire}. Cronbach alpha values per dimension were calculated (GS = 0.853, E = 0.907, SE = 0.881, P = 0.905) to test the inter-item reliability of each dimension. To categorise students into different levels of SRL competence, the K-means clustering \cite{macqueen_methods_1967} was performed separately for each group based on their scores on these dimensions. To maximise the average centroid distance with high interpretability of the clusters, two clusters (average centroid distance (control group) = -0.972, average centroid distance (intervention group) = -1.027) were applied: 1) high SRL cluster (control group: 25 students, intervention group: 27 students), 2) low SRL cluster (control group: 15 students, intervention group: 14 students). A Chi-square test was performed to identify whether there is a difference in the proportion of SRL levels across years.

\subsection*{Data Analysis}

In order to investigate students' engagement with their writing tasks, we have analysed their Google Docs log data using time series analysis. Time series is a sequence of time-ordered data. It is a prevalent approach to modelling complex behaviours in many disciplines and making predictions about future behaviours based on historic data including finance, engineering, and health sciences, but still uncommon in educational contexts \cite{shin_time_2017}. Time series can be decomposed into components for further inspection of certain behaviours. For example, the trend represents long-term movement in time series, and seasonality refers to a short-term periodic pattern under a fixed period. In this study, the seasonality component was used as a proxy to investigate the regularity in students' engagement with the reflective writing task. Unlike other learning activities, we asked students to reflect on the weekly contents with no obligation to perform the task at a specific time. Hence, how a particular student deliberately plans to work on a task at their own preferred time is related to their ability to regulate their learning activities. 

Seasonal decomposition from the \textit{Statsmodels} Python package\footnote{\url{https://www.statsmodels.org}} was applied. The additive seasonal decomposition was deployed due to a static seasonal component observed from the data \cite{hyndman_forecasting_2018}. The additive seasonal decomposition started by extracting the trend from the time-series data using the moving average method. The detrended data was further used to extract the recurring pattern, the seasonality. A fixed period of seven days was selected both for identifying trends through a 7-day moving average and as a model parameter to consider seasonality at a weekly interval. 
      
\begin{figure}[h!]
    \centering
    \includegraphics[width=\textwidth]{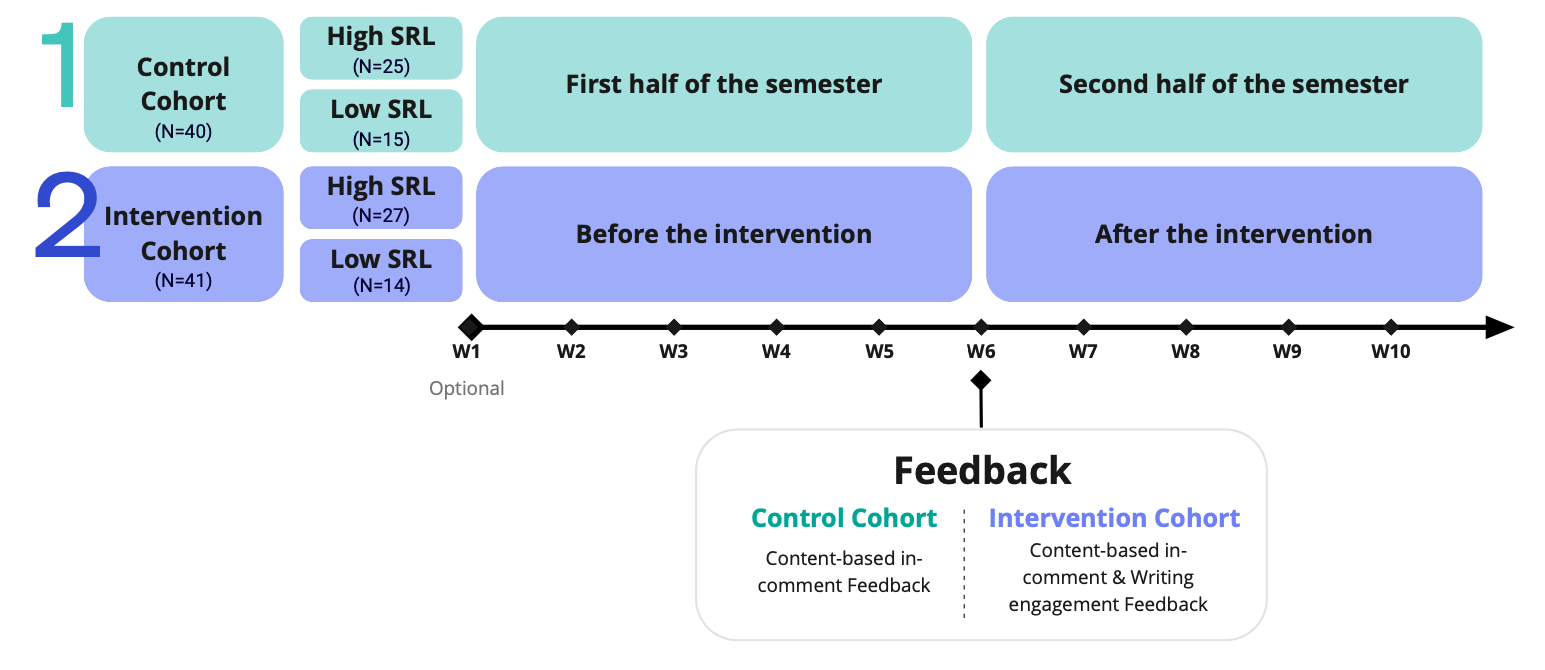}
    \caption{Overview of the study and feedback intervention across clusters of students}
    \label{fig:outline}
      \end{figure}

For comparative purposes, engagement data was divided into two periods: (1) before the study intervention (feedback email on students' engagement patterns) (Week 1 to 5) and (2) after the intervention until the end of the module (Week 6 to 10). Fig. \ref{fig:outline} represents overview of this study. In respect to the RQ1 and RQ2: how does the feedback intervention on students' reflective writing engagement affect their writing behaviours across years and across different levels of SRL, time series analysis and statistical comparison tests were applied to compare data at different levels: 1) at the cohort level (control and intervention groups) and 2) at the SRL cluster levels. Firstly, a time-indexed plot was used to visually represent an overview of the time series compared before and after the feedback. Then, seasonal decomposition was deployed to extract trends and seasonality. Secondly, descriptive statistics and statistical comparison tests were used to represent and confirm statistical differences across the groups. Regarding the RQ3, Pearson's r correlation was used to determine the relationship between students' final reflection scores and their derived writing engagement behaviours. In addition, an independent sample t-test was deployed to measure statistical significance differences in the derived writing engagement behaviours of the intervention cohort and the control cohort.

\section*{Results}

The subsequent sections represent year comparison, followed by varied SRL cluster comparisons using time series analysis and the results of the statistical comparison tests. The relationships between the derived writing engagement behaviours and the reflective scores of students are presented in the last section. 

\subsection*{1. Year comparison (\emph{RQ1: How does the intervention based on feedback about students’ reflective writing behaviours affect their engagement with the reflective writing task?})}
\subsubsection*{1.1 Time series analysis}

\begin{figure}
     \centering
     \begin{subfigure}[b]{0.8\textwidth}
         \centering
         \includegraphics[width=\textwidth]{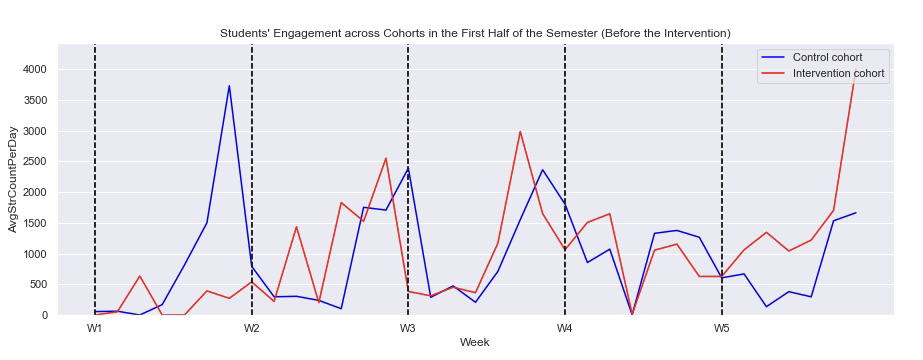}
         \caption{First half of the semester (before the intervention)}
         \label{fig:cohort comparison-before}
     \end{subfigure}
     \begin{subfigure}[b]{0.8\textwidth}
         \centering
         \includegraphics[width=\textwidth]{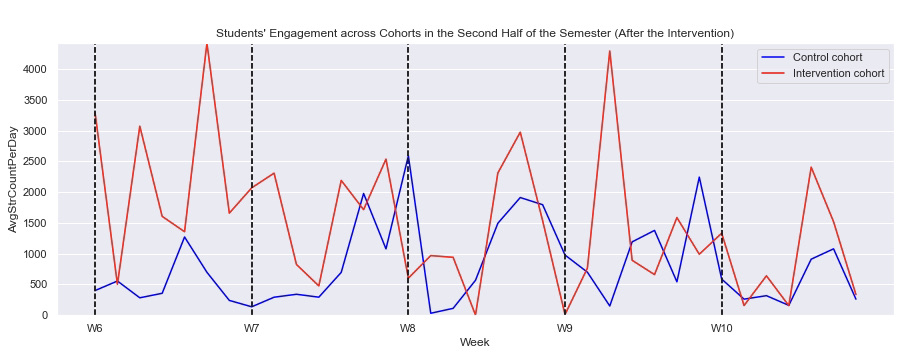}
         \caption{Second half of the semester (after the intervention)}
         \label{fig:cohort comparison-after}
     \end{subfigure}
        \caption{\emph{AvgStrCountPerDay} of both cohorts (the control group: blue line, the intervention group: red line) across semester. The vertical line denotes Monday of the week.}
        \label{fig:cohort comparison}
\end{figure}

Fig. \ref{fig:cohort comparison} shows the average number of edited strings per day (\emph{AvgStrCountPerDay}) of students compared between the control group (blue) and the intervention group (red) before (\ref{fig:cohort comparison-before}) and after the feedback intervention (\ref{fig:cohort comparison-after}). Both cohorts followed the same patterns of engagement before the intervention feedback (Fig \ref{fig:cohort comparison-before}), except in the first introductory week in which the control showed a surge in writing engagement although it was the optional week for reflection. On the other hand, whilst the control cohort exhibited similar visual patterns of engagement for the entire semester, the intervention cohort exhibited immediate and consistent engagement through higher numbers of \emph{AvgStrCountPerDay} after the intervention (Week 6).

\begin{figure}
    \addtocounter{subfigure}{0}
    \setcounter{subfigure}{0}
    \centering
     \begin{subfigure}[b]{0.8\textwidth}
        
        \centering
         \includegraphics[width=\textwidth]{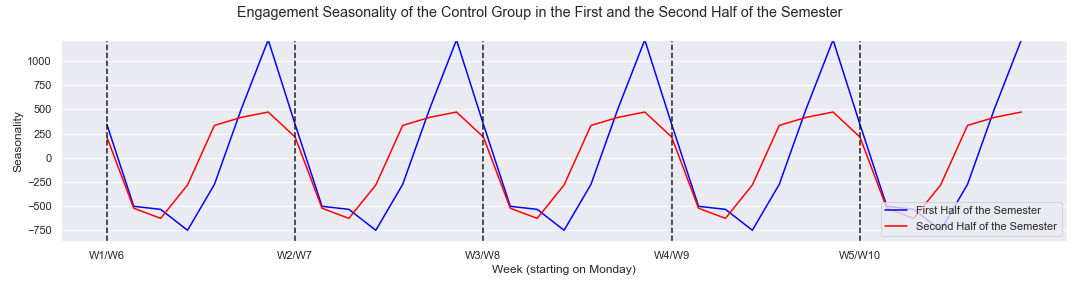}
         \caption{The control cohort}
         \label{fig:cohort comparison-seasonality-control}
     \end{subfigure}
     \begin{subfigure}[b]{0.8\textwidth}
        \centering
         \includegraphics[width=\textwidth]{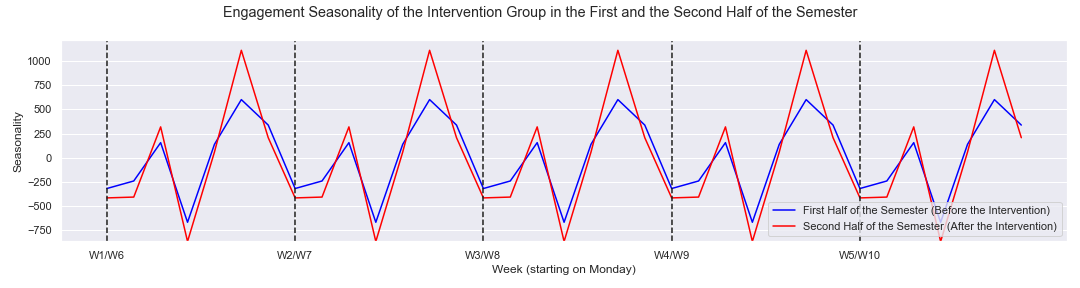}
         \caption{The intervention cohort}
         \label{fig:cohort comparison-seasonality-intervention}
     \end{subfigure}
        \caption{Extracted seasonality of \emph{AvgStrCountPerDay} of the control cohort (a) and the intervention cohort (b) compared between the first (blue) and the second half (red) of the semester (after the intervention).}
        \label{fig:cohort comparison-seasonality}
\end{figure}

7-day seasonality was extracted to observe weekly patterns to explore how students aligned their reflection behaviours in relation to other learning activities within the module. From Fig. \ref{fig:cohort comparison-seasonality}, there were observable visual differences in seasonality of \emph{AvgStrCountPerDay} between the control cohort (\ref{fig:cohort comparison-seasonality-control}, blue) and the intervention cohort (lower blue) before the feedback intervention. While the control cohort developed a 1-peak weekly pattern (peaks on Sundays), the intervention cohort demonstrated a 2-peak weekly pattern (peaks on Wednesdays and Saturdays) in which the seasonality of the cohorts highly deviated from each other. They both shared their weekly minima on Thursdays.
Similarly, 1-peak vs 2-peak weekly patterns among cohorts by calculating the number of editing frequencies per week among cohorts are presented in Table \ref{tab:cohort-freq}. The intervention cohort illustrated a higher percentage of editing twice or more times per week after the feedback intervention (from 28.29\% before the feedback to 32.20\% afterwards) whereas the control cohort showed a lower percentage of editing twice or more times per week.
\begin{table}[h!]
\centering
\caption{The number of editing frequency per week (\%) among cohorts}
\label{tab:cohort-freq}
      \begin{tabular}{lC{1.5cm}C{1.5cm}C{1.5cm}C{1.5cm}}
        \hline
        \multirow{2}{3cm}{\centering The number of editing frequency per week (\%)} & \multicolumn{2}{c}{Control cohort}  & \multicolumn{2}{c}{Intervention cohort}\\ 
        \cline{2-5}
         & Before the feedback & After the feedback & Before the feedback & After the feedback \\ \hline
        No edit & 28.00 & 45.00 & 40.49 & 35.12 \\
        Edit once per week & 47.50 & 35.50 & 31.21 & 32.68 \\
        Edit twice or more per week & 24.50 & 19.50 & 28.29 & 32.20 \\ \hline
      \end{tabular}
\end{table}

After the feedback, the intervention cohort expressed a consistent 2-peak engagement pattern at a weekly level and further intensified with a higher variance of engagement patterns. These can be recognised through the extreme deviation of the seasonality around zero (Fig. \ref{fig:cohort comparison-seasonality-intervention}, the red line). In contrast, earlier fall and rise in weekly engagement were spotted from the control cohort's seasonality on Wednesdays and Fridays respectively compared to Thursdays and Sundays in the period before the feedback (Fig. \ref{fig:cohort comparison-seasonality-control}, the red line). The differences in writing engagement patterns were further inspected using statistical comparison tests in the following section.

\subsubsection*{1.2 Statistical comparison tests}

There were four data sets namely daily engagement of (1) the control cohort in the first half of the semester, (2) the control cohort in the second half of the semester, (3) the intervention cohort in the first half of the semester (before the intervention), and (4) the intervention cohort in the second half of the semester (after the intervention), for comparison. Due to the non-normality of datasets, related-sample Wilcoxon signed-rank tests were used to compare differences in engagement before and after the feedback intervention within the same cohort. Additionally, Mann-Whitney U-tests were applied to identify differences in engagement behaviours between cohorts before and after the feedback intervention. 

Regarding differences within the same cohort, Wilcoxon signed-rank indicated that there was no statistically significant difference in \emph{AvgStrCountPerDay} of the control cohort between the first and the second half of the semester, \emph{Z} = -.508, \emph{p} = 0.612. In contrast, statistically significant difference in writing engagement behaviours of the intervention cohort between the first half (before the intervention) and the second half of the semester (after the intervention) was detected, \emph{Z} = -2.162, \emph{p} = 0.031, with a small effect size, \emph{r} = .26.  \emph{AvgStrCountPerDay} significantly increased from 1041.56 characters (IQR = 1193.24, N = 35) before the feedback intervention to 1354.17 characters (IQR = 1669.83, N = 35) after the feedback intervention (Table \ref{tab:cohort-stats1}). 
\begin{table}[h!]
    \centering
    \caption{Wilcoxon signed rank test on \emph{AvgStrCountPerDay} of the cohorts compared the first and the second half of the semester (before and after the intervention, respectively)}
    \label{tab:cohort-stats1}
      \begin{tabular}{C{2.5cm}C{0.8cm}C{1.5cm}C{0.8cm}C{1.5cm}C{0.8cm}C{0.8cm}C{0.7cm}}
        \hline
        \multirow{2}{*}{\emph{AvgStrCountPerDay}} & \multicolumn{2}{C{2.3cm}}{First half of the semester (Before the intervention)} & \multicolumn{2}{C{2.3cm}}{Second half of the semester 
(After the intervention)} & \multirow{2}{*}{\emph{Z}} & \multirow{2}{*}{\emph{p}} & \multirow{2}{0.7cm}{\centering Effect size}\\  
        \cline{2-5}
         & N (Inactive days) & Md (IQR) & N (Inactive days) & Md (IQR) \\ \hline
        Control Cohort & 35 (1) & 709 (1294.98) & 35 (0) & 564.35 (909.78) & -.508 & .612 & 0.06 \\
        Intervention Cohort & 35 (2) & 1041.56 (1193.24) & 34 (1) & 1354.17 (1669.83) & -2.162 & .031* & 0.26 \\ \hline
      \end{tabular}
      \raggedright
      *. \emph{p} is significant at the 0.05 level (2-tailed).\\
      IQR: interquartile range.
\end{table}

With respect to writing engagement differences across cohorts, Mann-Whitney U-test indicated that there was no statistically significant difference, \emph{U} = 593.00, \emph{p} = 0.819, in \emph{AvgStrCountPerDay} between the control (Md = 709, IQR = 1294.98) and the intervention cohort (Md = 1041.56, IQR = 1193.24) in the first half of the semester (before the intervention) (Table \ref{tab:cohort-stats2}). On the contrary, Mann-Whitney U-test demonstrated that there was significantly higher writing engagement in the intervention cohort (Md = 1354.17, IQR = 1669.83) compared to the control cohort (Md = 564.35, IQR = 909.78) after they received the feedback, \emph{U} = 369.00, \emph{p} = 0.004, with a medium effect size \emph{r} = .34.

\begin{table}[h!]
    \centering
    \caption{Mann-Whitney U-test of \emph{AvgStrCountPerDay} before and after the intervention compared between two cohorts}
    \label{tab:cohort-stats2}
      \begin{tabular}{L{2.5cm}C{0.8cm}C{1.3cm}C{0.8cm}C{1.3cm}C{0.9cm}C{0.9cm}C{0.7cm}}
        \hline
        \multirow{2}{2.5cm}{\emph{AvgStrCountPerDay}} & \multicolumn{2}{c}{Control cohort} & \multicolumn{2}{c}{Intervention cohort} & \multirow{2}{0.9cm}{\centering\emph{U}} &  \multirow{2}{0.9cm}{\centering\emph{p}} & \multirow{2}{0.7cm}{\centering Effect size}\\  
        \cline{2-5}
         & N (Inactive days) & Md (IQR) & N (Inactive days) & Md (IQR) \\ \hline
        First half of the semester (Before the intervention) & 35 (1) & 709 (1294.98) & 35 (2) & 1041.56 (1193.24) & 593.00 & .819 & 0.027\\
        Second half of the semester (After the intervention) & 35 (0) & 564.35 (909.78) & 34 (1) & 1354.17 (1669.83) & 369.00 & .004** & 0.34 \\ \hline
      \end{tabular}
      \raggedright
      **. \emph{p} is significant at the 0.01 level (2-tailed).\\
      IQR: interquartile range.
\end{table}

\subsection*{2. SRL Clusters' writing engagement results (\emph{RQ2: How does the feedback intervention affect the writing task engagement of students with different self-regulated learning (SRL) competence?})}

The following sections describe writing engagement results among high and low SRL groups across cohorts. It is important to note that the proportions of SRL levels did not differ by the two cohorts, \emph{x\textsuperscript{2}} (1, N = 81) = .099, \emph{p}  = .753.

\subsubsection*{2.1 Time series analysis}

\begin{figure}
     \centering
     \begin{subfigure}[b]{0.8\textwidth}
         \centering
         \includegraphics[width=\textwidth]{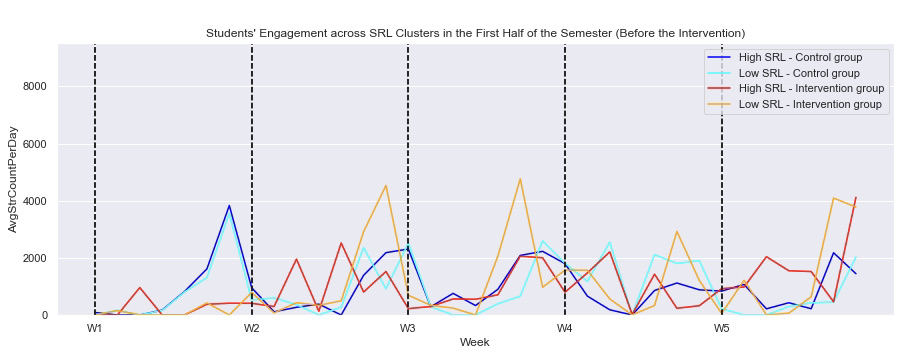}
         \caption{First half of the semester (before the intervention)}
         \label{fig:cluster comparison-before}
     \end{subfigure}
     \begin{subfigure}[b]{0.8\textwidth}
         \centering
         \includegraphics[width=\textwidth]{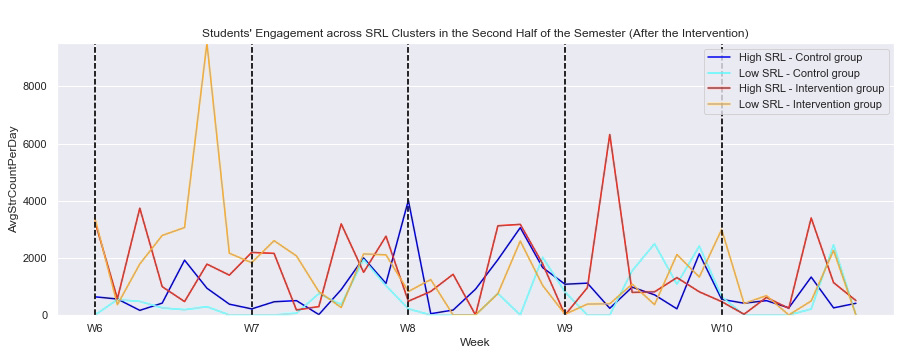}
         \caption{Second half of the semester (after the intervention)}
         \label{fig:cluster comparison-after}
     \end{subfigure}
        \caption{\emph{AvgStrCountPerDay} compared high SRL students from the control cohort (blue), low SRL students from the control cohort (cyan), high SRL students from  the intervention cohort (red), and low SRL students from the intervention cohort (orange) across semester.}
        \label{fig:cluster comparison}
\end{figure}

Time series plots for the first and the second half of the semester were depicted in Fig. \ref{fig:cluster comparison-before} and \ref{fig:cluster comparison-after}, respectively. Each plot shows a comparison in the number of \emph{AvgStrCountPerDay} across clusters: 1) the control cohort's high SRL score students (blue line), 2) the control cohort's low SRL score students (cyan line), 3) the intervention cohort's high SRL score students (red line), and 4) the intervention cohort's low SRL score students (orange line). Overall, there was a comparable amount of writing engagement from high and low SRL groups of the control cohort in the first half of the semester (Fig. \ref{fig:cluster comparison-before}, blue and cyan lines), yet higher engagement after first-half of the semester from the control cohort's high SRL group compared to the low SRL group of the same year was observed. There were lower numbers of inactive days (the day with no engagement) among groups in the second-half of the semester compared to the first-half of the semester, except in the control cohort's low SRL group (Table \ref{tab:cluster-stats}). 

\begin{table}[h!]
    \centering
    \caption{The number of editing frequency per week (\%) among clusters}
    \label{tab:cluster-freq}
      \begin{tabular}{L{2.5cm}C{.8cm}C{.8cm}C{.8cm}C{.8cm}C{.8cm}C{.8cm}C{.8cm}C{.8cm}}
        \hline
        \multirow{3}{2.5cm}{\centering The number of editing frequency per week (\%)} & \multicolumn{4}{c}{Control cohort} & \multicolumn{4}{c}{Intervention cohort} \\  
        \cline{2-9}
         & \multicolumn{2}{c}{High SRL} & \multicolumn{2}{c}{Low SRL} & \multicolumn{2}{c}{High SRL} & \multicolumn{2}{c}{Low SRL} \\ \cline{2-9}
         & H1 & H2 & H1 & H2 & H1 & H2 & H1 & H2 \\ \hline
        No edit & 31.20 & 42.40 & 22.67 & 49.33 & 38.52 & 37.78 & 44.29 & 30.00 \\
        Edit once per week & 40.80 & 32.80 & 58.67 & 40.00 & 30.37 & 31.85 & 32.86 & 34.29 \\
        Edit twice or more per week & 28.00 & 24.80 & 18.67 & 10.67 & 31.11 & 30.37 & 22.86 & 35.71 \\ \hline
      \end{tabular}
      \raggedright
      H1: First half of the semester (before the intervention) \\
      H2: Second half of the semester (after the intervention) \\
\end{table}

\begin{figure}
     \centering
     \begin{subfigure}[b]{0.8\textwidth}
         \centering
         \includegraphics[width=\textwidth]{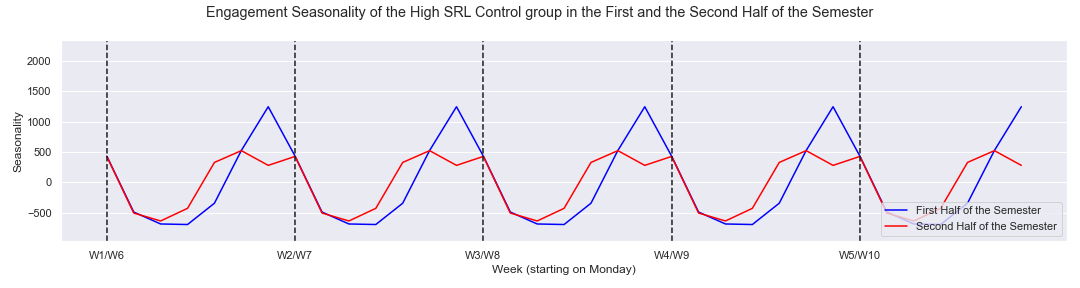}
         \caption{The control cohort's high SRL cluster}
         \label{fig:cluster comparison-seasonality-CH}
     \end{subfigure}
     \begin{subfigure}[b]{0.8\textwidth}
         \centering
         \includegraphics[width=\textwidth]{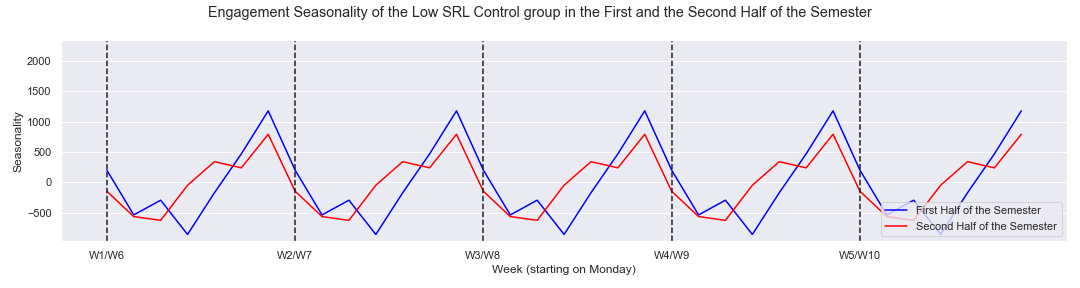}
         \caption{The control cohort's low SRL cluster}
         \label{fig:cluster comparison-seasonality-CL}
     \end{subfigure}
          \begin{subfigure}[b]{0.8\textwidth}
         \centering
         \includegraphics[width=\textwidth]{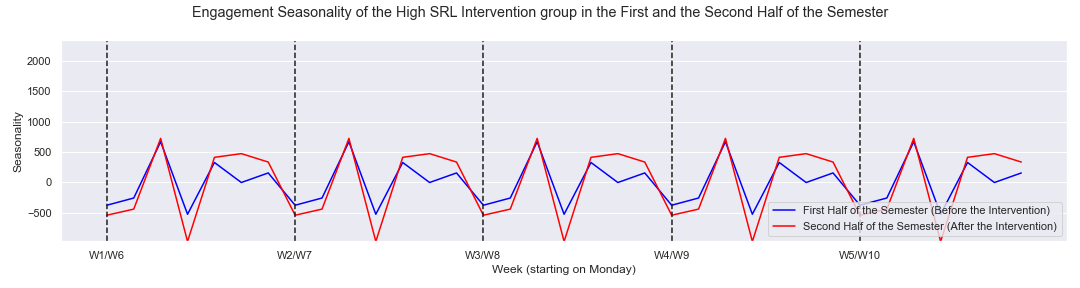}
         \caption{The intervention cohort's high SRL cluster}
         \label{fig:cluster comparison-seasonality-IH}
     \end{subfigure}
     \begin{subfigure}[b]{0.8\textwidth}
         \centering
         \includegraphics[width=\textwidth]{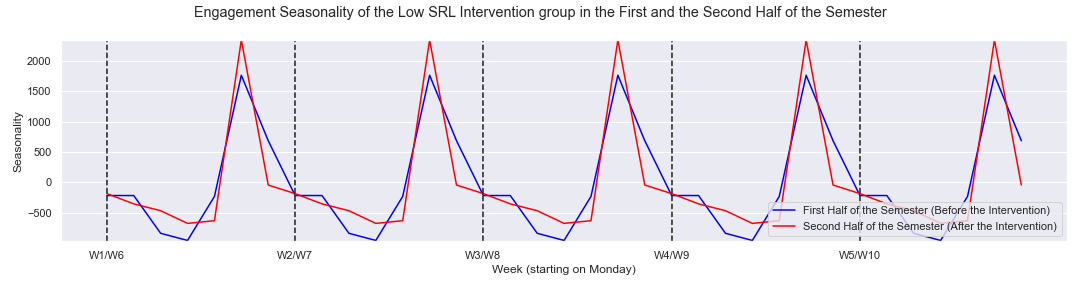}
         \caption{The intervention cohort's low SRL cluster}
         \label{fig:cluster comparison-seasonality-IL}
     \end{subfigure}
        \caption{Seasonal component of \emph{AvgStrCountPerDay} compared the first-half (before the intervention) (blue line) and the second-half of the semester (after the intervention) (red line) among clusters.}
        \label{fig:cluster comparison-seasonality}
\end{figure}

Regarding weekly patterns, seasonality compared across different SRL levels and cohorts was plotted in Fig. \ref{fig:cluster comparison}, where blue and red lines represented seasonality for the two halves of the semester, respectively. Fig. \ref{fig:cluster comparison-seasonality-CH}, \ref{fig:cluster comparison-seasonality-CL}, \ref{fig:cluster comparison-seasonality-IH}, \ref{fig:cluster comparison-seasonality-IL} represent engagement in the order of the control cohort's high SRL cluster, the control cohort's low SRL cluster, the intervention cohort's high SRL cluster, and the intervention cohort's low SRL cluster from the top to bottom accordingly. In general, every group illustrated homologous weekly patterns for both halves of the semester with minor changes. Considering the control cohort, corresponding seasonal patterns were observed in the high and low SRL groups for both periods  (Fig. \ref{fig:cluster comparison-seasonality-CH} and \ref{fig:cluster comparison-seasonality-CL}). To be specific, there was a 1-peak weekly pattern in which engagement amount was initially dropped from Mondays to reach its minimum on Thursdays and gradually increased towards the weekend, approaching its maximum amount on Sundays (blue lines) during the first half of the semester. These similar patterns were also observed after Week 5, yet higher nuances, i.e., multi-peaks were spotted from the seasonality patterns (orange lines). 

Instead, the high SRL group of the intervention cohort developed a distinctive 2-peak weekly engagement pattern for both before and after receiving the feedback intervention (Fig. \ref{fig:cluster comparison-seasonality-IH}). The pre- and post-intervention weekly patterns showed a sharp increase in engagement from Mondays to Wednesdays, followed by a significant drop on Thursdays and a continuing increase in engagement starting from Fridays to the weekend. Even though both seasonality from the pre- and post-feedback intervention periods followed a similar pattern, engagement after the feedback revealed relatively higher fluctuation in which the minimum and the maximums were respectively lower (Thursdays) and higher (Saturdays) compared to the period before the feedback intervention. Considering the weekly engagement pattern of the intervention's low SRL group, they demonstrated a 1-steep-peak pattern for both the periods before and after receiving the feedback intervention (Fig. \ref{fig:cluster comparison-seasonality-IL}). That is, there was a slight decrease pattern from Mondays to reaching the minimum on Thursdays, followed by a skyrocketed increase to the maximum on Saturdays before the feedback. However, after the feedback intervention, the minimum slightly shifted from Thursdays to Fridays whereas the maximum remained on the same day, yet higher variation was observed. Similar to the cohort level, the frequency of the edit per week among clusters was further analysed in Table \ref{tab:cluster-freq}. As anticipated, the high SRL group of the intervention cohort maintained their editing frequency twice or more times per week at approximately 30\% before and after the feedback. The low SRL group of the intervention cohort showed an increase in their editing frequency twice or more times per week from 22.86\% before the feedback to 35.71\% after the feedback. The following section represents statistical test results to investigate differences in \emph{AvgStrCountPerDay} within these four groups.

\subsubsection*{2.2 Statistical tests}

Similar to the cohort level, \emph{AvgStrCountPerDay} per cluster exhibited non-normal distribution; therefore, related-sample Wilcoxon signed-rank tests were applied for each cluster to identify within-sample differences in \emph{AvgStrCountPerDay} for each halves of the semester (Table \ref{tab:cluster-stats}). Only one test appeared to be significant, confirming statistical differences between \emph{AvgStrCountPerDay} of the intervention cohort's high SRL group before and after the feedback intervention, \emph{Z} = -2.244, \emph{p} = 0.025, with small effect size, \emph{r} = 0.27. In other words, there was a statistically significantly higher amount of \emph{AvgStrCountPerDay} before the feedback (Md = 705.81, IQR = 1225.56) compared to after the feedback (Md = 992.04, IQR = 1709.41). Apart from this group, no statistically significant difference was found in the amount of writing engagement within groups (High SRL students of the control cohort, \emph{Z} = -.328, \emph{p} = 0.743; Low SRL students of the control cohort, \emph{Z} = -1.820, \emph{p} = 0.069; Low SRL students of the intervention cohort, \emph{Z} = -.880, \emph{p} = 0.379). Despite observable visual variation in time series patterns before and after the feedback, only the high SRL group of the intervention cohort showed a statistically significant increase in their reflective writing engagement after receiving the feedback intervention. It is worth noting that increases in the writing engagement median were only observed in the intervention cohort's clusters.

\begin{table}[h!]
    \centering
    \caption{Wilcoxon signed rank test on \emph{AvgStrCountPerDay} of SRL clusters compared before and after the intervention}
    \label{tab:cluster-stats}
      \begin{tabular}{C{2.5cm}C{0.8cm}C{1.5cm}C{0.8cm}C{1.5cm}C{0.8cm}C{0.8cm}C{0.7cm}}
        \hline
        \multirow{2}{*}{\emph{AvgStrCountPerDay}} & \multicolumn{2}{C{2.3cm}}{First half of the semester (Before the intervention)} & \multicolumn{2}{C{2.3cm}}{Second half of the semester 
(After the intervention)} & \multirow{2}{*}{Z} & \multirow{2}{*}{p} & \multirow{2}{0.7cm}{\centering Effect size}\\  
        \cline{2-5}
         & N (Inactive days) & Md (IQR) & N (Inactive days) & Md (IQR) \\ \hline
        Control cohort \\
        \quad High SRL & 35 (4) & 822.44 (1228.36) & 35 (0) & 559.88 (863.36) & -.328 & .743 & 0.04 \\
        \quad Low SRL & 35 (6) & 457.87 (1649.73) & 35 (13) & 216.60 (805.93) & -1.820 & .069 & 0.22 \\ \hline
        Intervention cohort \\
        \quad High SRL & 35 (3) & 705.81 (1225.56) & 35 (2) & 992.04 (1709.41) & -2.244 & .025* & 0.27 \\
        \quad Low SRL & 35 (4) & 427.86 (1511.71) & 35 (4) & 1079.93 (1773.93) & -.880 & .379 & 0.11 \\ \hline
      \end{tabular}
      \raggedright
      *. \emph{p} is significant at the 0.05 level (2-tailed).\\
      IQR =  interquartile range.
\end{table}

\subsection*{3. Relationship between reflective writing engagement and academic performance (\emph{RQ3: What are the relationships between students’ writing engagement behaviours, and their final grades?})}

In this section, we present Pearson's r correlation results to identify the relationship between the engagement features and reflective scores, followed by independent sample t-test results comparing engagement features between control and intervention cohorts' first and second halves of the semester. 

\subsubsection*{3.1 Correlation between writing engagement features and reflective score}

From Table \ref{tab:correlation}, reflective score was found to be moderately positively correlated with TotalActiveWeek (\emph{r}(81) =.448, \emph{p} \textless .001) and weakly positively correlated with \emph{AvgStrCountPerDay} (\emph{r}(81) = .353, \emph{p} = 0.01) and \emph{TotalRevision} (\emph{r}(81) = .293, \emph{p} = .008). There were no significant correlations found between reflective score and other engagement features we derived namely \emph{AvgStrCountPerDay}, \emph{AvgRevPerDay}, \emph{AvgStrCountPerWeek} and \emph{AvgRevPerWeek}.

\begin{table}[h!]
    \centering
    \caption{Pearson correlations between reflective scores and the seven features}
    \label{tab:correlation}
      \begin{tabular}{L{2cm}C{1cm}C{1cm}C{1cm}C{1cm}C{1cm}C{1cm}C{1cm}}
        \hline
        & \emph{Total Rev} & \emph{AvgStr Count PerDay} & \emph{AvgRev PerDay} & \emph{Total Active Day} & \emph{AvgStr Count Per Week} & \emph{AvgRev Per Week} & \emph{Total Active Week} \\ \hline
        Reflective score & .293** & .186 & .210 & .353** & .183 & .152 &.448** \\
        \emph{p} & .008 & .097 & .060 & .001 & .101 & .174 & \textless.001 \\ \hline
      \end{tabular}
      \raggedright
      **. Correlation is significant at the 0.01 level (2-tailed).
\end{table}

\subsubsection*{3.2 Statistical tests on differences in writing engagement features before and after the feedback}

In the first half of the semester (Before the feedback period), there were no significant differences in the numbers of \emph{TotalRev}, \emph{t}(74) = -.711, \emph{p} = .480, \emph{TotalActiveDay}, \emph{t}(74) = -.479, \emph{p} = .633), and \emph{TotalActiveWeek} (\emph{t}(74) = 1.977, \emph{p} = .052) between two years (Table \ref{tab:t-test}). However, after the feedback, the intervention cohort demonstrated statistically significantly higher number of \emph{TotalActiveDay} (M = 6.05, SD =3.72) than the control cohort (M = 4.51, SD = 2.86), \emph{t}(74) = -2.012, \emph{p} = .048, whereas no other statistically significant differences were found in \emph{TotalRev}, \emph{t}(74) = -1.791, \emph{p} = .078, and \emph{TotalActiveWeek}, \emph{t}(74) = -1.494, \emph{p} = .139) across cohorts.

\begin{table}[h!]
    \centering
    \caption{Independent samples tests of writing engagement features between the first and the second half of the semester (before and after the intervention) compared between two cohorts}
    \label{tab:t-test}
      \begin{tabular}{L{3.5cm}C{1cm}C{1cm}C{1cm}C{1.2cm}C{1cm}C{1cm}}
        \hline
        \multirow{2}{3.5cm}{\centering Reflective writing features} & \multicolumn{2}{c}{Control cohort} & \multicolumn{2}{c}{Intervention cohort} & \multirow{2}{*}{\emph{t}(74)} & \multirow{2}{*}{\emph{p}} \\  
        \cline{2-5}
        & M & SD & M & SD \\ \hline
        \emph{First Half of the Semester (Before the intervention)} \\
        TotalRev & 5880.97 & 8155.11 & 7446.71 & 10860.03 & -.711 & .480 \\
        TotalActiveDay & 5.45 & 2.39 & 5.79 & 3.70 & -.479 & .633 \\
        TotalActiveWeek & 3.79 & 1.23 & 3.21 & 1.32 & 1.977 & .052 \\ \hline
        \emph{Second Half of the Semester (After the intervention)} \\
        TotalRev & 4438.81 & 7828.83 & 8826.87 & 13024.04 & -1.791 & .078 \\
        TotalActiveDay & 4.51 & 2.86 & 6.05 & 3.72 & -2.012 & \textbf{.048*} \\
        TotalActiveWeek & 2.97 & 1.26 & 3.41 & 1.29 & -1.494 & .139 \\ \hline
      \end{tabular}
      \raggedright
      *. \emph{p} is significant at the 0.05 level (2-tailed).\\
\end{table}

\section*{Discussion}
Although many previous studies attempted to provide analytics support on student writings, the value of content-specific analytics still appears to be limited in real-world settings, however, analytics of engagement with writing has the potential to provide value to students' reflective writing performance.  In this study, the effects of a feedback intervention that combines human educators' feedback on students' reflective writing content with the analytics of students' writing behaviours (intervention cohort), compared to human educators' content-only feedback (control cohort), were investigated. More specifically, the number of edited contents per day was selected as a proxy to represent students' engagement with a reflective writing task. Two cohorts' numbers of daily edited content data were examined using time series analysis to visually observe any potential pattern differences. Based on these observations, hypothesised differences between the intervention and control cohorts were tested using statistical comparison tests. Apart from cohort comparisons, the impact of the feedback interventions on students with varying degrees of SRL competence was further investigated.

With respect to RQ1: \emph{How does the intervention based on the feedback about students' reflective writing behaviours affect their engagement with the reflective writing task?} In line with our hypothesis, the results show that students who were provided with feedback on their writing behaviours engaged with the writing task significantly more after the combined feedback compared to the control cohort students. Despite the visually observed variation in weekly engagement patterns, both cohorts had similar amounts of engagement prior to the feedback intervention with no statistically significant difference. This suggests that presenting students with analytics of their writing engagement behaviours, in addition to educators' feedback on their writing content, can serve as an opportunity to encourage students' persistent engagement with their writing tasks. This result is aligned with previous research on the effects of behavioural feedback on engagement in other studies \cite{nelson_good_2012}. In addition, time series patterns demonstrated a surge in engagement immediately after the feedback intervention. This might indicate an increase in the responsiveness and timely reactions of students to the intervention provided. Such characteristics of students are shown to correlate with a higher ability to regulate learning and higher academic performance \cite{suraworachet_examining_2021}. Moreover, the seasonality of the two cohorts revealed anticipated engagement patterns in which the lowest interaction was spotted on Thursdays (the day when there was demand from another module in the same programme) and the highest engagement was reached during the weekend. However, these weekly patterns varied between the control and intervention cohorts in which the intervention group exhibited a bimodal weekly engagement pattern compared to the unimodal pattern in the control group. A bimodal seasonality coupling with an increase in weekly engagement frequency may refer to a higher frequency of reflection per week. It may represent students' strategy to divide the tasks into smaller units for better coping with tasks and/or an engagement pattern beyond fulfilling the minimum of the module requirement \cite{fredricks_school_2004} which are associated with better learning outcomes \cite{yip_learning_2012}. However, these interpretations require further qualitative in-depth investigations of students' experiences to be confirmed.

Regarding the RQ2: \emph{How does the feedback intervention affect the writing task engagement of students with different SRL competence?} The low and high SRL groups of the intervention cohort showed increasing engagement with the writing tasks after receiving the engagement feedback as evident through their daily writing behaviours and the similar number of active days before and after the intervention. Conversely, the low and high SRL groups of the control cohort exhibited lower engagement in the second half of the module (Week 6-10) in which the control cohort's low SRL group distinctively showed a higher number of inactive days afterwards. These withdrawal effects after the sole content-based feedback in low SRL' s control cohort are aligned with Mitchell, McMillan and Rabbani's \cite{mitchell_exploration_2019} study which showed that low self-efficacy students reported higher negative feelings or anxiety emerging from the content feedback alone. Moreover, the low SRL group of the intervention cohort was the only group that showed higher percentages of engagement with the writing task with twice or more times engagement within a week after the feedback intervention provided. These results indicate the positive impact of the writing analytics feedback intervention on students' engagement with the writing task mainly occurs through changing the routine behaviours of low SRL students while not influencing the engagement regularity of high SRL students significantly. These results further support the previous research evidence \cite{nelson_good_2012} that low-performing students might benefit even more from timely feedback on their engagement behaviours. In terms of engagement quantity, despite variations in engagement patterns compared before and after the intervention, the intervention cohort's low SRL group did not show statistically significant differences in their daily engagement quantity. On the contrary, the intervention cohort's high SRL group was the only group that showed statistically significant higher engagement quantity after the intervention compared to the period before. Thus, these results indicate that the analytics feedback intervention is likely to help students with low SRL competence recognise the necessity of regular reflections, maintain their engagement with the reflective writing tasks and help them improve their academic performance while showing no detrimental effects on high SRL students.

Going back to our third research question: \emph{What are the relationships between students' reflective writing behaviours, and their academic performance?}, significant correlations between students' final grades on their reflective writing tasks and writing engagement behaviours were observed. These indicated that the quantitative features of reflective writing behaviours concerning the quantity of students' writing (both at the daily and the weekly level) were not correlated to their final grades. Instead, there were weak to moderate size positive correlations between the students' final reflective writing  scores and the regularity of their engagement with the writing tasks. This confirms previous research that the quality of students' individual writing is more related to the regularity of their reflections rather than the amount of reflective writing they produce in crammed sessions \cite{suraworachet_examining_2021}. The value of spaced practice \cite{sobel_spacing_2011, rohrer_effects_2006} and interleaving \cite{rohrer_interleaving_2012} for students have also been shown in multiple other studies from the learning sciences literature \cite{carpenter_spacing_2014}. We further investigated the extent to which the writing engagement behaviours changed in the feedback intervention cohort. The results showed that there was a higher regularity (measured through the observation of total active days in the reflective writing days) in students' engagement with the individual reflective writing task at a daily level in the intervention cohort after the feedback compared to the control cohort.

\subsection*{Limitation and Future Research}

Finally, several limitations should be noted. First, this is not a randomised controlled trial study in which the true effects of the feedback intervention can be claimed in causal arguments. Due to the ethical and practical concerns of studying real-world teaching and learning contexts, true randomisation of the intervention group and the control group was not possible. Therefore, we opted in for a quasi-experimental set-up in which two different cohorts were used as intervention and control groups. Although the engagement behaviours of two cohorts were observed before the intervention and no statistically significant differences were detected, the contextual variations of two different cohorts might influence the results presented in this study. For this reason, and also due to the limited sample sizes in both cohorts, it is difficult to generalise the results into other contexts without further investigations. Hence, this study encourages similar future work to be conducted in the field to expand understanding of the proposed feedback intervention's effect on students' writing engagement behaviours in other contexts. Moreover, it is worth highlighting that although we used a particular group of writing behaviours as proxies to predict students' writing engagement, there are other possible proxies which might be equivalently worth exploring (i.e, average revisions, time spent, the number of writing sessions, etc.). In addition to the current methods of analysis, an additional study on students' opinions on the feedback intervention and an extended analysis of the content presented could help provide a more comprehensive picture of students' understanding of the feedback provided and its intentional impact on their engagement with the writing task at behavioural, cognitive and emotional levels.

\section*{Conclusion}

This quasi-experimental study demonstrates a robust investigation of the real-world effects of an original human educator and analytics combined feedback intervention on students' writing engagement behaviours in an authentic semester-long graduate module. The intervention group consisted of students who received personalised engagement feedback with analytics during the mid-term in addition to their content feedback from educators and exhibited higher engagement during the second half of the semester after receiving the analytics feedback. Additionally, they also demonstrated higher regularity in engaging with the reflective writing task both at a daily and weekly level which significantly positively correlated with their academic performance. The combined feedback intervention was found to be more effective, especially for students with low SRL competence. This result contributes to the broader research in reflective writing support to consider coupling feedback from both cognitive and behavioural aspects to match learners' SRL levels. It is particularly significant given the context independency of the behavioural feature we engineered and the ubiquitous use of the Google Docs platform for generating the analytics feedback. However, further investigations on the longevity of the impacts, as well as their cross-context validity, should be undertaken. 


\begin{backmatter}

\section*{Abbreviations}
SRL: Self-Regulated Learning, NLP: Natural Language Processing, RQ: Research Question, GS: Goal-Setting, E: Effort, SE: Self-Efficacy, P: Persistence

\section*{Availability of data and materials}
The datasets used and analyzed during the current study are available from the corresponding author on reasonable request.

\section*{Funding}
Not applicable

\section*{Ethics approval and consent to participate}
Ethical approval was received from the institutional ethics review committee. 

\section*{Acknowledgements}
We would like to thank DUTE 2020/2021 and 2021/2022 students for granting permissions to collect data for this study and also thank Prof. Yannis Dimitriadis for his valuable comments on an earlier version of the manuscript.

\section*{Competing interests}
The authors declare that they have no competing interests.


\section*{Authors' contributions}
    All authors contributed in constructing the materials, intervention design for the study, data analysis and paper writing. All authors read and approved the final manuscript.

\section*{Authors' information}
UCL Knowledge Lab, Institute of Education, University College London, 23-29 Emerald St, London, UK.\\ Wannapon Suraworachet, Qi Zhou and Mutlu Cukurova

\printbibliography
\end{backmatter}
\newpage
\section*{Appendix}
\subsection*{Appendix A: Email feedback}
\label{appendix:email}
\begin{table}[H]
    \centering
    \caption*{Sample email of the designed engagement feedback}
    \begin{tabular}{|p{12cm}|}
      \hline
\\Dear ,\\
\\
First of all, thank you for continuing to engage with the individual reflection. You must have already received feedback on the content of your reflections as comments in the shared document. This email is a formative evaluation of your individual reflection behaviours. We hope this can help you to better regulate your reflective writing behaviours. \\
\\
According to previous research on students' reflective writing behaviours, high self-regulated learners tend to 1) show systematic patterns by allocating a specific time slot for completing the reflection task, 2) react promptly to the feedback and 3) avoid postponing the work until the deadline. More information could be found here:\textit{anonymised} 
\\
\\ 
The graph attached here shows a comparison of the number of edited strings per day across five weeks between you (red), last year's cohort (blue) and this year's cohort (green). Compared to last year's cohort, your cohort are more active in the first 5 weeks. Well done for all your hard work! \\
\\
In the first 5 weeks, you have been active for 3 weeks. Although there was a delay during the first few weeks, we are glad to see you became more active by mid-term. According to the graph, you have completed most of your writing in Week 5. It would be better to allocate specific time to complete the individual reflection. Writing regularly after reading or discussing may help with your individual reflection.\\
\\
It is worth highlighting that the higher the number of edited contents itself might not always lead to better learning outcomes but the systematic pattern of reflective writing behaviours can.\\
\\
The analytic data presented here are for reflective purposes, not part of your summative assessment.
We look forward to reading more of your individual reflections. \\
\\
Please let us know if you have any questions or concerns.\\
\\
Best regards\\\\

\hline
    \end{tabular}
\end{table}

\subsection*{Appendix B: SRL questionnaire}
\label{appendix:questionnaire}
\begin{longtable}[c]{|L{0.6cm}|L{2cm}|L{4.2cm}|L{4cm}|}
\caption*{Questions and references\label{long}}\\
 \hline
 No. & Category & Question & References\\ \hline \endfirsthead
 \hline
 No. & Category & Question & References\\ \hline \endhead
 \hline
 \endfoot
    1	&	Goal Setting	&	I set standards for my assignments in a class/subject/module.	&	OSLQ (Barnard et al., 2009)	\\
    2	&	Goal Setting	&	I set short-term (daily or weekly) goals as well as long-term goals (monthly or for the semester).	&	OSLQ (Barnard et al., 2009)	\\
    3	&	Goal Setting	&	I keep a high standard for my learning in a class/subject/module.	&	OSLQ (Barnard et al., 2009)	\\
    4	&	Goal Setting	&	I set goals to help me manage study time for a class/subject/module.	&	OSLQ (Barnard et al., 2009)	\\
    5	&	Persistence	&	Regardless of whether or not I like materials in a class/subject/module, I work my hardest to learn it.	&	Persistence (Elliot et al., 1999)	\\
    6	&	Persistence	&	When something that I am studying gets difficult, I spend extra time and effort trying to understand it.	&	Persistence (Elliot et al., 1999)	\\
    7	&	Persistence	&	I try to learn all of the testable material "inside and out,'' even if it is boring.	&	Persistence (Elliot et al., 1999)	\\
    8	&	Persistence	&	I work hard to do well in a class/subject/module even if I don't like what we are doing.	&	Effort regulation (Pintrich et al., 1991)	\\
    9	&	Persistence	&	Even when class/subject/module materials are dull and uninteresting, I manage to keep working until I finish.	&	Effort regulation (Pintrich et al., 1991)	\\
    10	&	Persistence	&	When I was feeling bored, I forced myself to pay attention. &	Motivation control (Warr \& Downing, 2000)	\\
    11	&	Persistence	&	When my mind began to wander during a learning session, I made a special effort to keep concentrating.	&	Motivation control (Warr \& Downing, 2000)	\\
    12	&	Persistence	&	I increased my effort when the material did not really interest me.	&	Motivation control (Warr \& Downing, 2000)	\\
    13	&	Persistence	&	I pushed myself even harder when I began to lose interest. &	Motivation control (Warr \& Downing, 2000)	\\
    14	&	Persistence	&	Whenever I lost interest in my work, I made a special effort to pay attention.	&	Motivation control (Warr \& Downing, 2000)	\\
    15	&	Effort	&	I usually spent more time than the requirements of my class/subject/module.	&	Effort (Adapted from Fisher \& Ford, 1998)	\\
    16	&	Effort	&	I usually provide extra effort in my class/subject/module.	&	Time on task (Adapted from Brown, 2001)	\\
    17	&	Self-efficacy	&	I'm certain I can understand the basic concepts in any class/subject/module.	&	Self-efficacy for learning and performance (Pintrich et al., 1991)	\\
    18	&	Self-efficacy	&	I believe I will receive an excellent grade in any class/subject/module.	&	Self-efficacy for learning and performance (Pintrich et al., 1991)	\\
    19	&	Self-efficacy	&	I'm certain I can understand the most difficult material presented in the readings for any class/subject/module.	&	Self-efficacy for learning and performance (Pintrich et al., 1991)	\\
    20	&	Self-efficacy	&	I'm confident I can learn the basic concepts taught in any class/subject/module.	&	Self-efficacy for learning and performance (Pintrich et al., 1991)	\\
    21	&	Self-efficacy	&	I'm confident I can understand the most complex material presented by the instructor in any class/subject/module.	&	Self-efficacy for learning and performance (Pintrich et al., 1991)	\\
    22	&	Self-efficacy	&	I'm confident I can do an excellent job on the assignments in any class/subject/module.	&	Self-efficacy for learning and performance (Pintrich et al., 1991)	\\
    23	&	Self-efficacy	&	I expect to do well in any class/subject/module.	&	Self-efficacy for learning and performance (Pintrich et al., 1991)	\\
    24	&	Self-efficacy	&	I'm certain I can master the skills being taught in any class/subject/module.	&	Self-efficacy for learning and performance (Pintrich et al., 1991)	\\
    25	&	Self-efficacy	&	Considering the difficulty of this module, the teacher, and my skills, I think I will do well in any class/subject/module.	&	Self-efficacy for learning and performance (Pintrich et al., 1991)	\\  \hline
\end{longtable}

\end{document}